\documentclass[a4paper]{IEEEtran}
\usepackage[utf8]{inputenc} 
\usepackage{amsmath}
\usepackage{amssymb}
\usepackage{graphicx}
\usepackage{hyperref}
\usepackage{mathtools}
\usepackage{changepage} 
\usepackage{multirow}
\usepackage{float}
\usepackage{authblk}
\usepackage[table,xcdraw]{xcolor}

\usepackage{makecell}

\usepackage{mathptmx}
\usepackage{setspace}

\usepackage{etoolbox}

\title{An RNA Sequencing Analysis of Glaucoma Genesis in Mice}
\author[1]{Jai Sharma}
\author[2,*]{Vidhyacharan Bhaskar}
\affil[1]{Student Intern Researcher, Robinson Lab, Stanford University; \newline 269 Campus Drive, Palo Alto, CA 94305; Email: jaisharmaus@gmail.com}
\affil[2]{Professor, School of Engineering, San Francisco State University; \newline 1600 Holloway Avenue, San Francisco, CA 94132}
\affil[*]{Corresponding Author's Email: vcharan@gmail.com}

\begin{document}
\maketitle
\begin{abstract}
Glaucoma is the leading cause of irreversible blindness in people over the age of 60, accounting for 6.6 to 8\% of all blindness in 2010, but there is still much to be learned about the genetic origins of the eye disease. With the modern development of Next-Generation Sequencing (NGS) technologies, scientists are starting to learn more about the genetic origins of Glaucoma. This research uses differential expression (DE) and gene ontology (GO) analyses to study the genetic differences between mice with severe Glaucoma and multiple control groups. Optical nerve head (ONH) and retina data samples of genome-wide RNA expression from NCBI (NIH) are used for pairwise comparison experimentation. In addition, principal component analysis (PCA) and dispersion visualization methods are employed to perform quality control tests of the sequenced data. Genes with skewed gene counts are also identified, as they may be marker genes for a particular severity of Glaucoma. The gene ontologies found in this experiment support existing knowledge of Glaucoma genesis, providing confidence that the results were valid. Future researchers can thoroughly study the gene lists generated by the DE and GO analyses to find potential activator or protector genes for Glaucoma in mice to develop drug treatments or gene therapies to slow or stop the progression of the disease. The overall goal is that in the future, such treatments can be made for humans as well to improve the quality of life for human patients with Glaucoma and reduce Glaucoma blindness rates.
\end{abstract}


\section{Introduction} 

\subsection{Background}
\indent The neurodegenerative eye disease Glaucoma is a part of a group of eye diseases which can cause blindness. Glaucoma in particlar is one of the leading causes of blindness in people over the age of 60 \cite{glaucoma_info}. The most common type of Glaucoma, Primary Open Angle Glaucoma (POAG)\footnote{This paper will be discussing only Primary Open Angle Glaucoma, so the terms ``Glaucoma'' and ``Primary Open Angle Glaucoma'' will be used interchangeably unless otherwise specified.}, is caused when resistance in the Trabecular Meshwork leads to increased Intraocular Pressure (IOP) in the eye. As a result, aqueous humor, which supplies nutrients to the cornea, is unable to flow from the Posterior Chamber to the Anterior Chamber, through the Trabecular Meshwork, and back into the body. In POAG, a gradual increase in IOP leads to damage to the optic nerve and the chronic and irreversible onset of Glaucoma. As POAG progresses, the IOP increase leads to cupping of the optic disc; while the ratio between the diameter of the optic cup to the diameter of the optic disc is 0.5 in a normal eye, this ratio increases during cupping in POAG \cite{glaucoma_info}. Left untreated, Glaucoma takes on average 25 years to progress from normal vision to complete blindness. Even with treatment, 9\% to 19\% of patients reach complete blindness, implying a need for treatment improvements \cite{thesis}. 
\\
\\
\indent Glaucoma in general is characterized by the cupping and atrophy of the optic nerve head, visual field loss of the patient, and often the increased IOP \cite{who}. Thapa, \textit{et al.} studied a subset of the Nepalese population and found that the occurrence rate of Glaucoma was 1.9\%. Approximately 68\% of these cases were Primary Open Angle Glaucoma, 22.67\% of these cases were Primary Angle Closure Glaucoma, and 9.33\% of these cases were secondary Glaucoma \cite{nepal}. Casson, \textit{et al.} highlighted that while Glaucoma used to be characterized by high IOP, many cases of Glaucoma without high IOP and high IOP without Glaucoma disputed this assumption. Now, high IOP is simply considered a risk factor rather than a defining feature, and reducing IOP is currently the only treatment strategy for Glaucoma \cite{concepts}.

\subsection{Risk Factors and Symptoms}
\indent It is known that old age, family history, Black ethnic origin, and Myopia (nearsightedness) are all risk factors for Glaucoma. Once a patient is diagnosed with Glaucoma, routine screening by an optometrist is used to ensure the disease's slow progression. As POAG progresses, the patient experiences a loss of peripheral vision (tunnel vision), fluctuating pain, headaches, and blurred vision. Symptoms are usually worse at nighttime or in dark spaces \cite{glaucoma_info}.

\subsection{Current Diagnosis and Treatment Methods}
\indent While 111.8 million people are expected to have Glaucoma by 2040, there is much to be learned about the genetic origins and development of this disease. Glaucoma (all types collectively) have a 10\% diagnosis rate worldwide and less than 50\% diagnosis rate in developed nations \cite{thesis}. 
\\
\\
\indent Currently, there are several ways to diagnose Glaucoma. First, Non-Contact Tonometry can be used to shoot puffs of air and measure puff response from the cornea. A small reaction to the puff of air would be associated with high IOP, and an according Glaucoma diagnosis can be made. This method is relatively less accurate, but it is a good estimation for eye pressure. The industrial standard for Glaucoma diagnosis is the Goldmann Applanation Tonometry method, which utilizes advanced equipment to make contact with the cornea and apply different amounts of pressure to find the IOP, from which a Glaucoma diagnosis can be made. Other forms of diagnosis include Fundoscopy, which can help examine the eye for cupping, and the Visual Field Assessment Test, which tests for peripheral vision loss \cite{glaucoma_info}. Glaucoma is asymptomatic until it reaches an advanced stage, so it is recommended that patients undergo regular screening examinations from age 40 onward. The rate of false positives for Glaucoma diagnosis is high, so it is also recommended that any positive finding is followed up with further testing \cite{diagnosis}. 
\\
\\
\indent The IOP for a normal eye ranges from 10 mmHg to 21 mmHg, and treatment is often started when IOP reaches 24 mmHg. Current treatments include Prostaglandin Analogues (e.g. Latanoprost) which increase uveoscleral outflow, Topical Beta-Blockers (e.g. Timolol) and Carbonic Anhydrase Inhibitors (e.g. Dorzolamide) which reduce the production of aqueous humor, and Sympathomimetics (e.g. Brimonidine) which increase uveoscleral outflow and reduce the production of aqueous humor. Surgical methods of treatment include Trabeculectomy which creates a new channel for aqueous humor to drain from the eye; however, this procedure is risky and its effects are only temporary \cite{glaucoma_info}. Conlon, \textit{et al.} analyzed trends in Glaucoma treatments, concluding that there has been a recent increase in Glaucoma medications and the use of laser trabeculoplasty over the past decade. Additionally, there has been a noticable decrease in the frequency of invasive incisional surgery. New methods such as microinvasive Glaucoma surgery have also been introduced \cite{trends}. Despite advances in laser and incisional surgery, Schwartz, \textit{et al.} claim that medical therapy is still the primary method of treatment and that there is good evidence that laser trabeculoplasty is just as good as initial medical therapy \cite{management}.

\subsection{Research Goal}
\indent Crucial to the development of new cures for POAG is the more sophisticated understanding of the disease at the genetic level. The goal of this research is to find the genes responsible for the genesis of Glaucoma in mice so that this information may be used to develop new treatments in the future which target these genes or corresponding gene pathways in mice. The overall goal is that future research can develop new and more effective POAG treatments for humans.
\\
\\
\indent This research is focused on generating candidate gene lists which may contain potential activator or protector genes for Glaucoma in mice. However, it should be noted that doing intensive biological studies to study these gene lists manually is outside the scope of this research.
\\
\\
\indent Note that mice were used here because mice have eyes which are similar to human eyes in that they develop hereditary Glaucoma and high IOP, as will be explained in the next section. This research and similar research on Glaucoma have taken advantage of this unique property of mice.

\section{Literature Review}
\indent Research has been conducted to explore the function of ncRNA, transcriptional mRNA, and similar biological molecules in Glaucoma, Cataracts, and different types of cancers. This research has shown to successfully identify targets for further therapeutic research to develop treatments for the corresponding disease. 
\\
\\
\indent 
Chen, \textit{et al.} explored time-series circRNA, lncRNA, miRNA, and mRNA expression profiles of developing mice retina samples. The aim of this research was to find the key functional ncRNA and ceRNA which regulated retinal neurogenesis. The research found that several ncRNAs in the circRNA/lncRNA-miRNA-mRNA network, including circCDYL, circATXN1, circDYM, circPRGRIP, lncRNA Meg3, and lncRNA Vax2os, involved neurotransmitter transport and multicellular organism growth during retinal development \cite{bmc}. Zhou, \textit{et al.} aimed to identify differing biological mechanisms involved in each Glaucoma subtype. Results from the research indicated that human participants with high tension Glaucoma had an enrichment of genes associated with unfolded protein response. Further, the research identified the differential expression of genes in certain IOP regulating tissues \cite{thesis}. Howell, \textit{et al.} used house mice mRNA data set to investigate the genetic origins of Glaucoma using temporally ordered sequences of Glaucoma states. The overall goal was to identify molecular events which can be targeted to provide effective new treatments for human glaucoma. A major finding from this study was that mice with a mutation in the C1QA gene were protected from Glaucoma. Additionally, it was found that inhibition of endothelin system with bosentan, an endothelin receptor antagonist, was strongly protective against damage from Glaucoma \cite{ncbi_paper}. 
\\
\\
\indent Current research methods perform genetic comparisons between mice with Glaucoma and mice without Glaucoma, but one of the drawbacks of this research is that it fails to stratify these comparison based on specific features such as age or strain. Thus, we lack insight into how these features specifically contribute to or protect from Glaucoma. The research presented in this paper intends to solve this issue.
\\
\\
\indent This research focuses on finding potential activator or protector genes for Primary Open Angle Glaucoma in house mice. This research is significant because one control group of mice, matched by age but not strain, interestingly does not develop Glaucoma. It is worthwhile to examine the genetic differential expression between this control group and the severe Glaucoma group to better understand on the genetic level how one strain is apparently ``immune'' to Glaucoma. Other control groups used in this research consists of mice which are matched by age and strain but do not have Glaucoma and mice which are matched by strain and not age. Through these experiments, we can more effectively perform genetic comparisons to understand the genes which play a role in the genesis of Glaucoma in each mentioned case.

\section{Data Availability Statement}
\indent This research used a publicly available data set from NCBI (NIH) of RNA Sequence samples from the optical nerve head (ONH) and retina of mice \cite{ncbi_data} (citation is clickable link to data source). The following section is a description of the data set.
\\
\\
\indent Genome-wide assessment of gene expression was performed in two species of mice: DBA/2J mice and D2-Gpnmb+ mice. DBA/2J is a breed of mice which develops high IOP and hereditary Glaucoma similar to that in human eyes \cite{species_1}. D2-Gpnmb+, on the other hand, is a breed of mice which does not develop high IOP or Glaucoma \cite{species_2}. As such, D2-Gpnmb+ mice can be used as a control group when studying the DBA/2J mice. 
\\
\\
\indent The NCBI data set consisted of 40 RNA Seq samples from DBA/2J mice, each of 10.5 months of age, placed into one of the following groups based on Glaucoma severity: 1) No or early 1, 2) No or early 2, 3) Moderate, and 4) Severe.
\\
\\
\indent Additionally, the data set contained two control groups. One control group (Control Group 1) consisted of 10 D2-Gpnmb+ mice, each of 10.5 months of age (age and strain matched, no glaucoma control). The second control group (Control Group 2) consisted of 10 DBA/2J mice, each of 4.5 months of age (young, pre-glaucoma). Only ONH, not retina, RNA Seq data was provided for the second control group.

\section{Proposed Experimental Design}
\subsection{Methodology Description}
\indent An initial quality control analysis was used to gauge the usability of the data. Principal Component Analysis and Dispersion Analysis were used as initial visualizations. After this, genes with skewed expression were identified in hopes of finding marker genes from Glaucoma.
\\
\\
\indent After quality control, differential expression and gene ontology analyses were used to find potential activator or protector genes for Glaucoma.

\subsection{Experimental Design}
\[\includegraphics[scale=0.27]{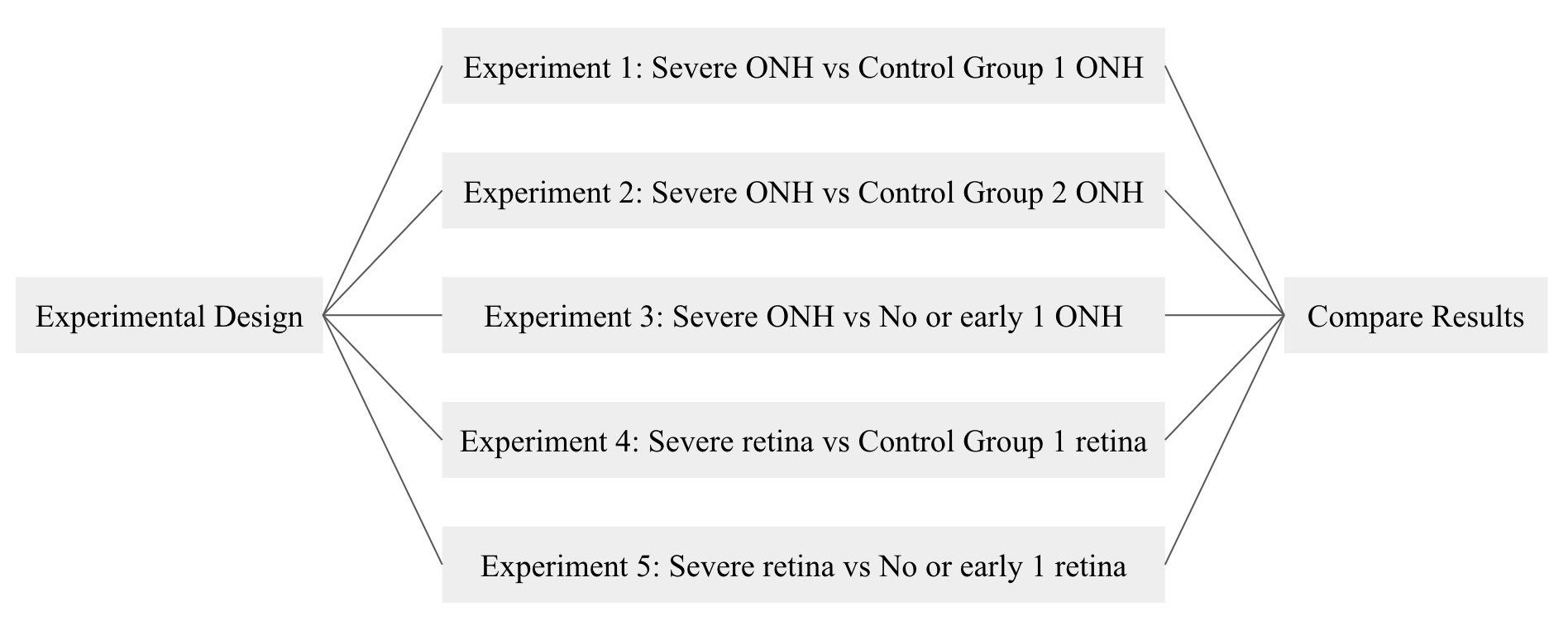}\]
\begin{align*}
    &\textbf{Figure 1: }\text{Experimental Design clarifying the comparisons}\\
    &\text{which were made in this experiment.}
\end{align*}
\indent The workflow of the comparisons outlined in Figure 1 will be clarified in the next section. Here, severe ONH and retina data are compared to multiple ONH and retina data control groups. The results of these comparisons are compared at the end of the study. In this study, the results of the different pairwise comparisons yielded similar differentially expressed genes and gene ontologies, giving more confidence in these results.
\\
\\
\indent To find the genes responsible for the genesis of Glaucoma, the Severe group was compared to three other groups: the two control groups and the No or early 1 group. The following is the intuition behind these comparisons:
\\
\\
\indent The comparison between the Severe and the Control 1 group could reveal which differentially expressed (DE) genes enable the D2-Gpnmb+ breed to not develop Glaucoma or high IOP. The comparison between the Severe and the Control 2 group could reveal which DE genes lead to the manifestation of Glaucoma as the mice grow older. The comparison between the Severe and the No or early 1 group could reveal which DE genes lead some mice to develop Glaucoma and not others (even when they are the same species and age). The No or early 1 group acted as a sort of third control group because the mice in this group do not develop Glaucoma.
\\
\\
\indent Additionally, the experiment was repeated for both the ONH and retina samples (where data was available).

\subsection{Experimental Workflow of Comparisons}
\indent This section clarifies the exact workflow of each experiment highlighted in the last section.
\\
\\
\indent Each workflow consisted of four main parts:
\begin{enumerate}
    \item Find differentially expressed (DE) genes between the two RNA Seq sample groups. Split this list of DE genes into an upregulated and downregulated gene list according to the value of the Log Fold Change (LogFC).
    \item Make heatmaps to compare the expression of the top 20 DE genes across all considered samples in the comparison. The top 20 DE genes consisted of the top 10 upregulated DE genes and the top 10 downregulated DE genes (ranked by LogFC).
    \item Make boxplots which model the expression of the top 20 upregulated DE genes and the the top 20 downregulated DE genes. The $x$-axis had each of the eleven RNA Seq sample groups in the data:
    \\
    \\
    No or early 1 ONH, No or early 2 ONH, Moderate ONH, Severe ONH, Control Group 1 ONH, Control Group 2 ONH, No or early 1 retina, No or early 2 retina, Moderate retina, Severe retina, and Control Group 1 retina.
    \\
    \\
    The $y$-axis represented the expression (in Log Gene Counts) of the gene. Since each boxplot visualization represented the expression of one DE gene, 40 boxplot visualizations were made per comparison. Recall that there were repeated boxplot visualizations made since a gene that was differentially expressed in one comparison could have been differentially expressed in another.
    \item Gene ontology (GO) was performed on the top 1000 upregulated DE genes (ranked by LogFC) and the Biological Process (BP), Cellular Component (CC), and Molecular Function (MF) gene ontologies were found. The gene ontology was similarly performed on the top 1000 downregulated DE genes. The DAVID gene ontology tool was used to conduct the gene ontology analysis \cite{david_1}\cite{david_2}.
\end{enumerate}

The results of Step 4 provide the more general result as to which group or class of genes should be targeted to create a new Glaucoma treatment for the DBA/2J mice.

\subsection{Other Analysis and Visualizations}
\indent In addition to the comparisons already mentioned, several other analyses were conducted.
\\
\\
\indent First, the Principal Component Analysis (PCA) algorithm was used to visualize the relationships between the different RNA Seq sample groups.
\\
\\
\indent Second, a Dispersion Visualization was made to assess the quality of the data and find the common biological coefficient of variation (BCV). 
\\
\\
\indent Third, the skewness in gene counts for each gene was calculated. Genes which were considered to have a skewness in their expression were visualized through histograms. Genes with skewed expression are important because they can serve as marker genes for a particular RNA Seq sample group.


\section{Experimental Results and Analysis}
\subsection{PCA Visualizations}
\indent The Principal Component Analysis (PCA) algorithm was used to visualize the optical nerve head (ONH\footnote{In the data set, optical nerve head is abbreviated as OHN, so the abbreviation OHN is used in the visualizations. However, these two abbreviations are essentially the same otherwise.}) and retina data. Since these two types of data come from different parts of the eye, they were visualized separately.
\[\includegraphics[scale=0.15]{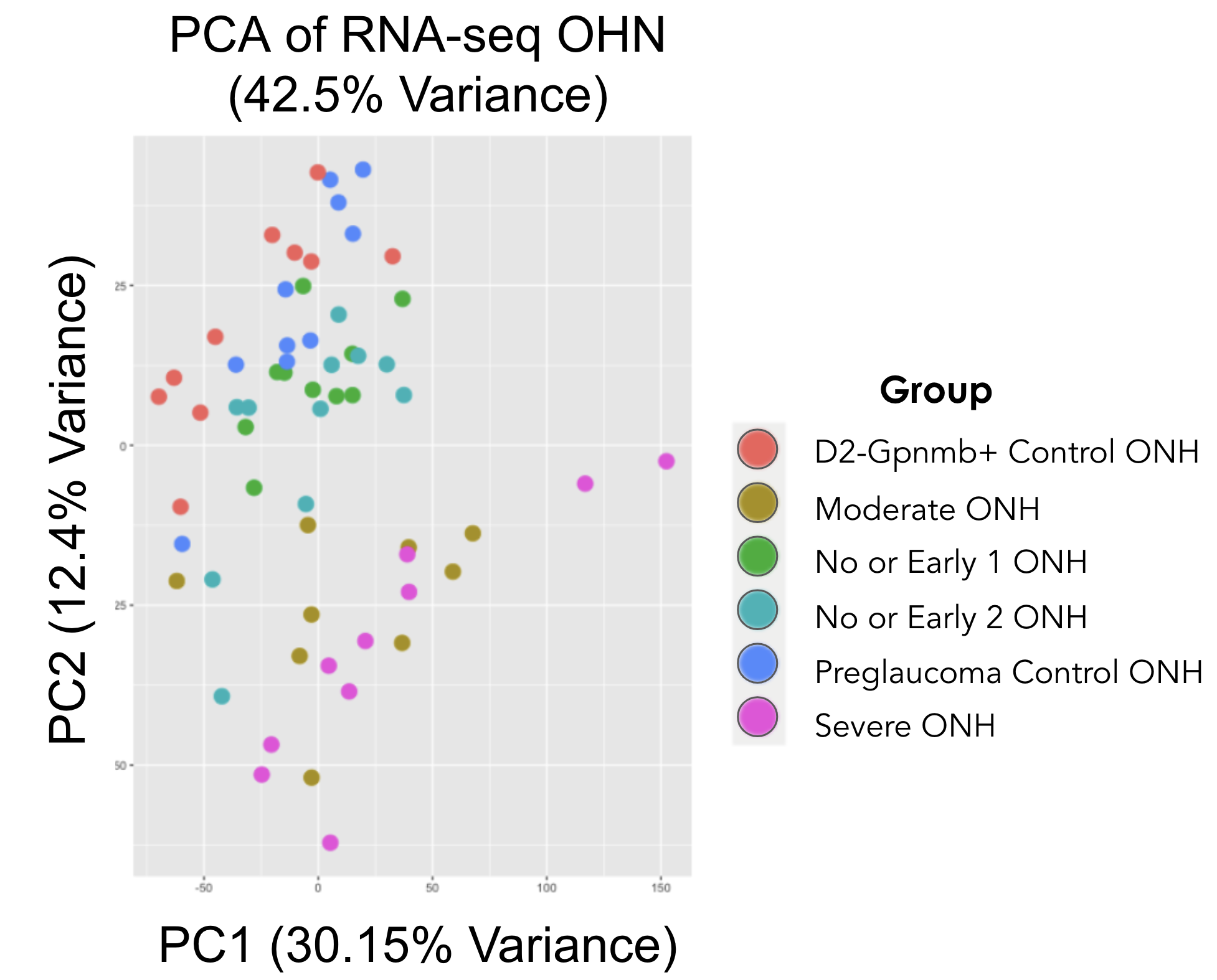}\]
\begin{align*}
    &\textbf{Figure 2: }\text{The PCA visualization of the ONH data using}\\
    &\text{principal components 1 and 2.}
\end{align*}
\indent As shown in Figure 2, groups Control Group 1 ONH, Control Group 2 ONH, No or early 1 ONH, and No or early 2 ONH all coincide with each other. This made sense due to the similarity of their nature; these were either control groups or groups which represent mice which do not develop Glaucoma as they age. On the other hand, the Moderate OHN and Severe OHN groups also coincided because they represent a different class of mice which do develop Glaucoma as they age.
\[\includegraphics[scale=0.15]{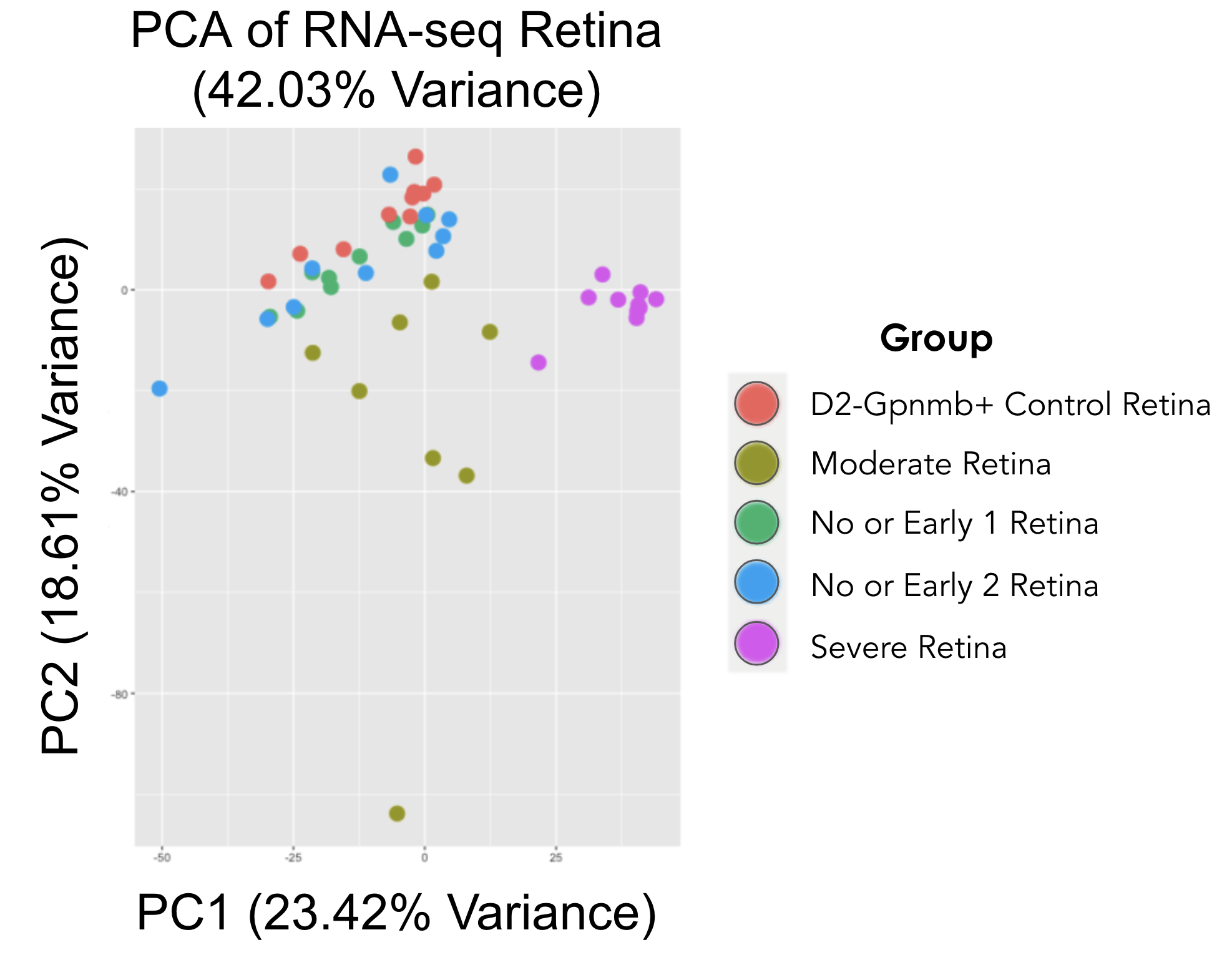}\]
\begin{align*}
    &\textbf{Figure 3: }\text{The PCA visualization of the retina data using}\\
    &\text{principal components 1 and 2.}
\end{align*}

Similar to the previous visualization, Figure 3 shows that groups Control Group 1 retina, Control Group 2 retina, No or early 1 retina, and No or early 2 retina all coincide with each other, while the Moderate retina and Severe retina groups are separate from the other groups.
\\
\\
\indent Note that the variance maintained in these visualizations (42.55\% and 42.03\%) is not especially high, so there are nuances in the data that cannot be seen here. 
\\
\\
\indent First, the fact that the Severe group, Moderate group, and other groups were separated in the visualization provided a good quality control check for the data set. 
\\
\\
\indent Furthermore, the stark contrast between the Severe group and the other groups implies that there may be significant genetic differences in these groups. Learning and targeting these genetic differences can aid the development of future treatments for Glaucoma. This reasoning served as a motivation for differential expression analysis and gene ontology.

\subsection{Dispersion Visualization}
\indent The dispersion plot visualization was used to visualize the biological coefficient of variation (BCV) for the data.
\[\includegraphics[scale=0.22]{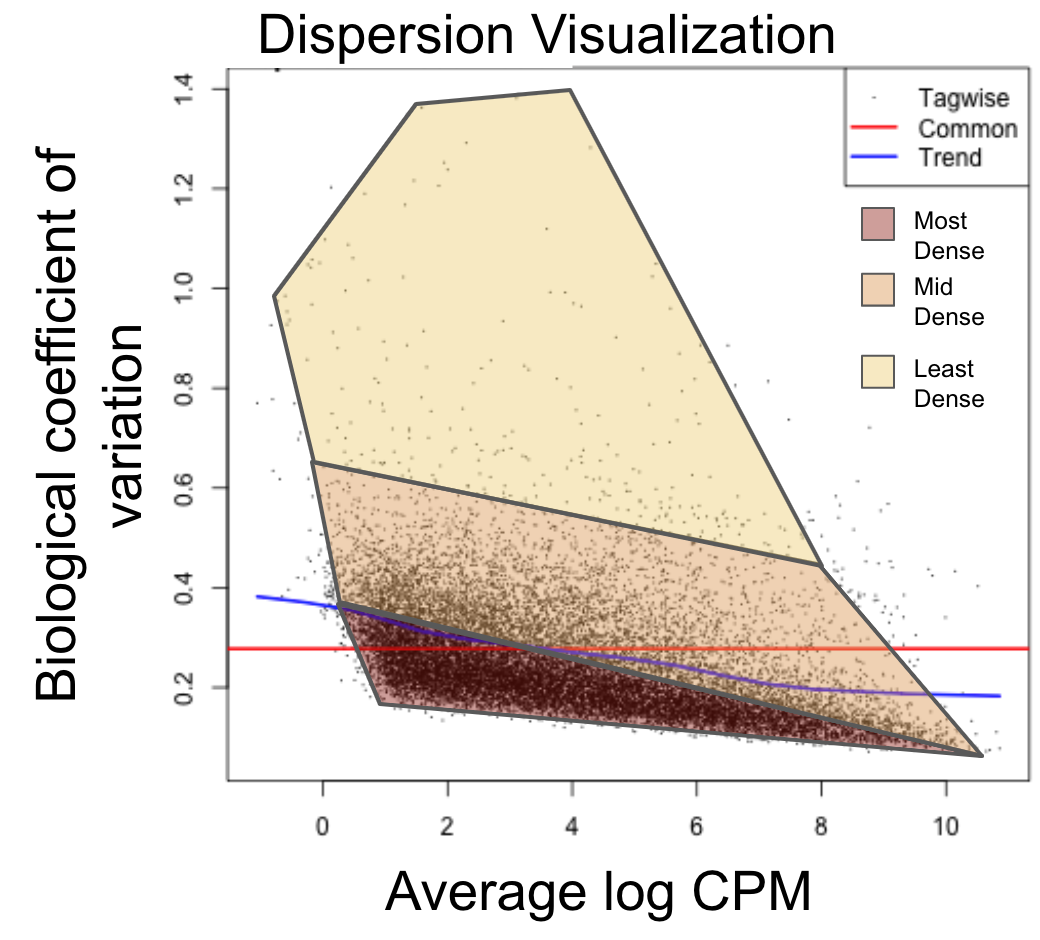}\]
\begin{align*}
    &\textbf{Figure 4: }\text{The dispersion visualization of the RNA Seq}\\
    &\text{data to visualize the BCV.}
\end{align*}
\indent In Figure 4, the Dispersion Visualization is shown. Each point corresponds to one gene, and the $x$ axis represents the average log (base 2) counts per million (CPM) of that gene while the $y$ axis represents the biological coefficient of variation (BCV) for that gene. The red line shows the common, or average, BCV, and the blue line shows the trend of the BCV as the average log CPM increases. 
\\
\\
\indent The common dispersion is $7.71\times 10^{-2}$, so the common BCV is $2.78\times 10^{-1}$. The industrial standard of BCV is less than about $3.00\times 10^{-1}$, so the common BCV of $2.78\times 10^{-1}$ was an acceptable level of variation. Additionally, the most dense area of points is almost entirely below the common BCV, the mid dense area of points is almost cut in half by the common BCV, and the least dense area of points is entirely above the common BCV. This means that the majority of genes have a BCV less than the common BCV, but there are certain genes which skew this plot and increase the common BCV. Nevertheless, the common BCV falls under the industrial threshold as mentioned previously, so this dispersion plot provided another good quality check of the data. These quality control procedures enabled the conduction of other analyses, as will be discussed in the following sections.

\subsection{Skewed Gene Counts}
\indent The Pearson Mode Skewness was used to determine the skewness in the gene counts of each gene across all data samples. The threshold 2 was used to determine which genes should be considered skewed in their gene counts. Specifically, if the Pearson Mode Skewness was less than -2 or greater than 2, a gene was considered skewed in gene counts.
\\
\\
\indent By this method, 112 genes were found to have skewed gene counts. For each selected gene, a histogram was constructed of the log gene counts of that gene across all samples. Specifically, the $x$-axis displayed the log gene counts, and the $y$-axis displayed the number of samples, or frequency.
\[\includegraphics[scale=0.3]{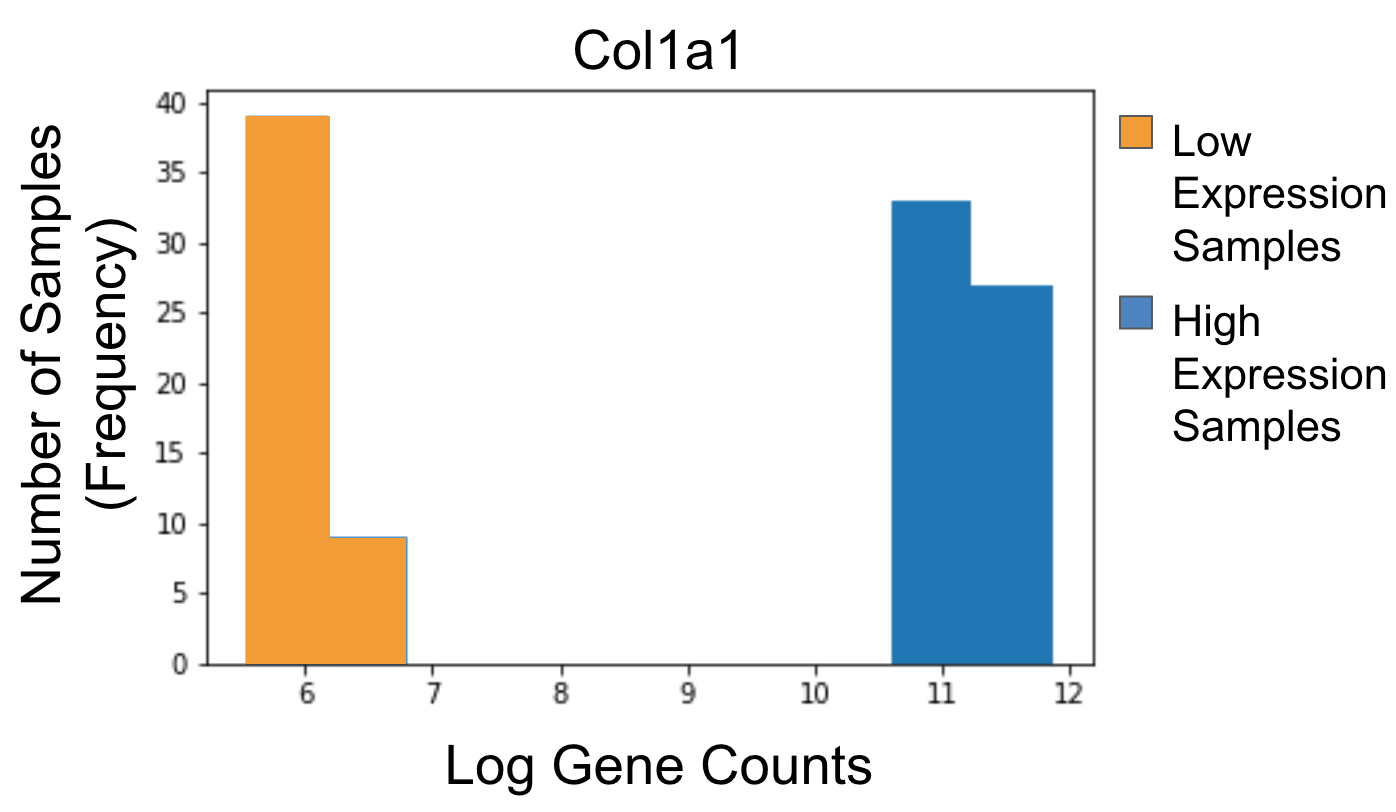}\]
\begin{align*}
    &\textbf{Figure 5: }\text{The histogram for the gene COL1A1 (Collagen }\\
    &\text{Type I Alpha 1 Chain), a gene which was identified to}\\
    &\text{be skewed in gene counts.}
\end{align*}
\indent As seen in Figure 5, the bimodal distribution implied that this gene could be used as a marker gene for some of the RNA Seq groups. Similarly, other genes found to have skewed expression were highlighted because they could potentially be marker genes for Glaucoma. Therefore, instead of having to perform whole genome sequencing, future researchers can focus on these marker genes for more efficient diagnosis of Glaucoma. 
\\
\\
\indent Not all 112 histograms could be shown here, so the rest of the histograms were uploaded to the GitHub repository for this paper \cite{github}.
\subsection{Pairwise Comparisons}
The complete results for the five pairwise comparisons were not included, but an example of the heatmap generated from the first pairwise comparison is shown below. The complete generated data is available on the GitHub repository for this paper. \cite{github}.


\paragraph{Heatmap} $\hspace{10mm}$
\[\includegraphics[scale=0.25]{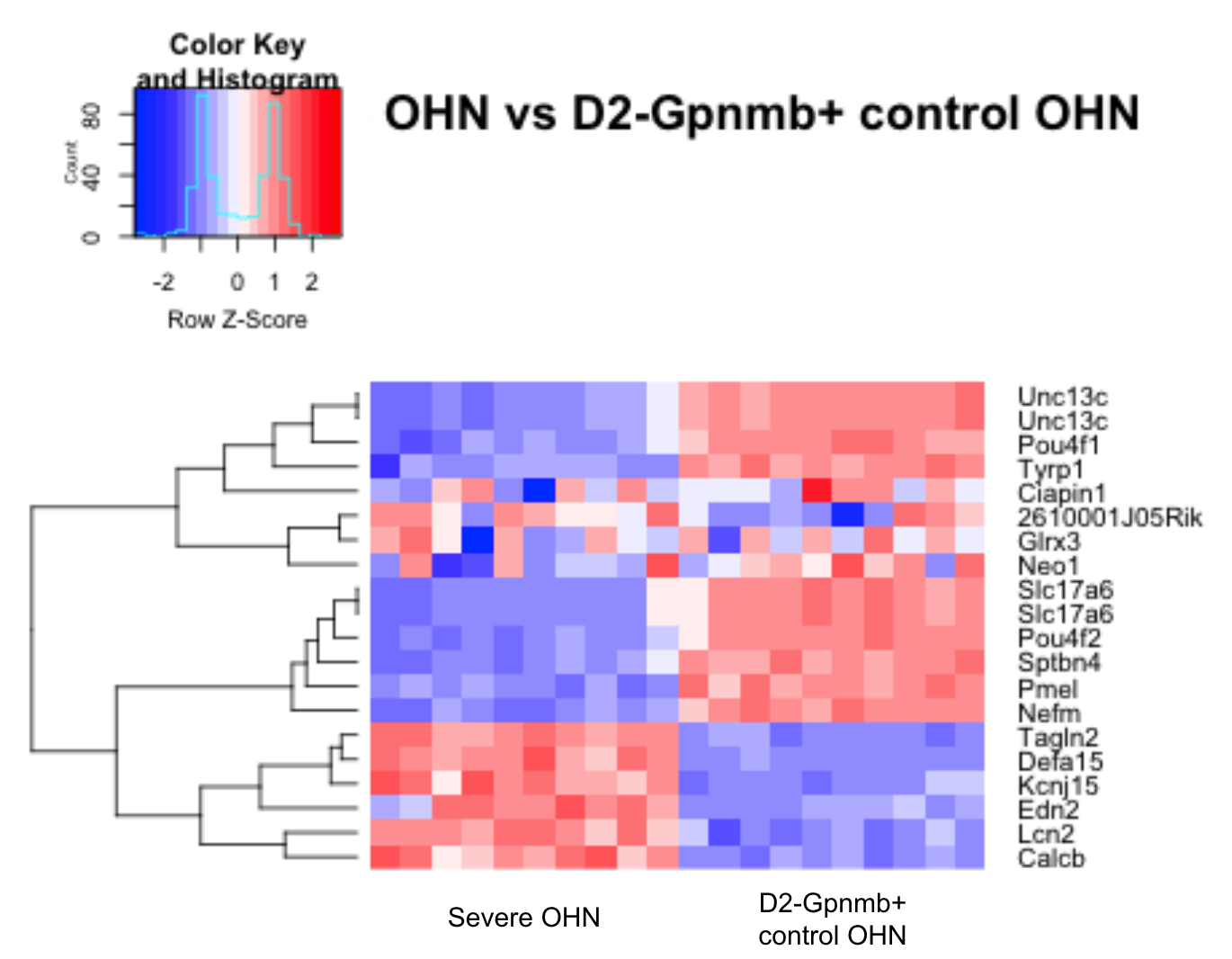}\]
\begin{align*}
    &\textbf{Figure 6: }\text{Heatmap of top 20 DE genes; Experiment 1: }\\
    &\text{Severe ONH vs Control Group 1 ONH.}
\end{align*}

\section{Discussions}
\indent This Discussion section discusses the conclusions, key takeaways, and implications of the pairwise comparison experiments.
\subsection{Conclusions and Key Takeaways}
\indent First, it should be noted that apart from several genes, the majority of the heatmaps contained genes which had clear differential expression between the two groups compared in each comparison. The stark contrast in the expression for the DE genes implied that the quality of the data was good, as we would expect such contrasts when comparing mice with severe Glaucoma and control groups.
\\
\\
\indent The key takeaways from this study lie in the gene ontologies. Specifically, common gene ontologies were found using the upregulated genes from different pairwise comparison experiments, implying that the results found were significant. These gene ontologies also support existing knowledge about the development of Glaucoma, implying that the results were valid.
\\
\\
\indent In biological process (BP) gene ontologies, nervous system development and synaptic signaling were common themes. Correspondingly, cellular component (CC) gene ontologies in locations such as the axon, synaptic membrane, and neuronal cell body were observed. Finally, these gene ontologies corroborated the molecular function (MF) gene ontologies, which included ion channel activity and gated channel activity.
\\
\\
\indent For each gene ontology, specific upregulated and downregulated DE gene lists were generated which correspond to that gene ontology. These gene lists, which can be found on the GitHub repository for this paper \cite{github}, can be studied and then selectively targeted in mice with Glaucoma to develop more effective treatments for the disease.
\\
\\
\indent It should be noted that this Discussion section does not go into a deep biological study of the individual genes in the generated gene lists as this is beyond the scope of this research; this goal is addressed in detail to future researchers in the next section.

\subsection{Need For Further Research}
\indent The differential expression and gene ontology analyses from the pairwise comparison experiments provide candidate gene lists which must be targeted to develop new Glaucoma treatments. Future researchers can perform thorough biology-oriented studies of these candidate gene lists to find potential activator or protector genes for Glaucoma in mice. These findings can be confirmed by inhibiting or activating such genes in mice to observe the effect on the development of Glaucoma or high IOP. 
\\
\\
\indent In accordance with these results, future researchers may develop new gene therapies or drugs which can be tested on mice to slow or stop the progression of Glaucoma. 
\\
\\
\indent Once such treatments are shown to be effective on mice, the possibility of creating treatments for human Glaucoma can be explored. Future researchers must also keep in mind the ethical concerns with human trials and how human based treatment testing may be different than mouse based treatment testing.

\end{document}